\documentclass[sigconf]{acmart}
\usepackage{booktabs} % For formal tables
\usepackage{multirow}
\usepackage{colortbl}
\usepackage{balance}
\newcommand{\includevisio}[2][]{\includegraphics[clip, trim=0.5cm 0.5cm 0.5cm 0.5cm, #1]{#2}} 

\usepackage{array}
\newcommand{\PreserveBackslash}[1]{\let\temp=\\#1\let\\=\temp}
\newcolumntype{C}[1]{>{\PreserveBackslash\centering}p{#1}}
\newcolumntype{R}[1]{>{\PreserveBackslash\raggedleft}p{#1}}
\newcolumntype{L}[1]{>{\PreserveBackslash\raggedright}p{#1}}

\setcopyright{none}
\renewcommand\footnotetextcopyrightpermission[1]{}
\newcommand\blfootnote[1]{%
 \begingroup
 \renewcommand\thefootnote{}\footnote{#1}%
 \addtocounter{footnote}{-1}%
 \endgroup
}

\acmDOI{}

\acmISBN{}

\acmConference[OSIRRC]{OSIRRC 2019 co-located with SIGIR 2019}{25 July 2019}{Paris, France}
\acmYear{2019}
\copyrightyear{2019}

\begin{document}

\title{Let's measure run time!} 

\subtitle{Extending the IR replicability infrastructure to include performance aspects}

\author{Sebastian Hofst{\"a}tter}
\affiliation{%
  \institution{TU Wien}
}
\email{s.hofstaetter@tuwien.ac.at}

\author{Allan Hanbury}
\affiliation{%
  \institution{TU Wien}
}
\email{hanbury@ifs.tuwien.ac.at}

\begin{abstract}

Establishing a docker-based replicability infrastructure offers the community a great opportunity: measuring the run time of information retrieval systems. The time required to present query results to a user is paramount to the users satisfaction. Recent advances in neural IR re-ranking models put the issue of query latency at the forefront. They bring a complex trade-off between performance and effectiveness based on a myriad of factors: the choice of encoding model, network architecture, hardware acceleration and many others. The best performing models (currently using the BERT transformer model) run orders of magnitude more slowly than simpler architectures. We aim to broaden the focus of the neural IR community to include performance considerations -- to sustain the practical applicability of our innovations. In this position paper we supply our argument with a case study exploring the performance of different neural re-ranking models. Finally, we propose to extend the OSIRRC docker-based replicability infrastructure with two performance focused benchmark scenarios. 

\end{abstract}

\maketitle
\blfootnote{Copyright \textcopyright\ 2019 for this paper by its
authors. Use permitted under Creative Commons License Attribution 4.0
International (CC BY 4.0). OSIRRC 2019 co-located with SIGIR 2019, 25
July 2019, Paris, France.}

\section{Introduction}

The replicability and subsequent fair comparison of results in Information Retrieval (IR) is a fundamentally important goal. Currently, the main focus of the community is on the effectiveness results of IR models. We argue that in the future the same infrastructure supporting effectiveness replicability should be used to measure \textit{performance}. We use the term \textit{performance} in this paper in the sense of speed and run time -- for the quality of retrieval results we use \textit{effectiveness}. In many cases, the time required to present query results to a user is paramount to the users satisfaction, although in some tasks users might be willing to wait longer for better results \cite{teevan2013slow}. Thus a discussion in the community about existing trade-offs between performance and effectiveness is needed. 

This is not a new insight and we don't claim to re-invent the wheel with this position paper, rather we want to draw attention to this issue as it becomes more prevalent in recent advances in neural network methods for IR. Neural IR ranking models are re-rankers using the content text of a given query and a document to assign a relevance score. Here, the choice of architecture and encoding model offer large effectiveness gains, while at the same time potentially impacting the speed of training and inference by orders of magnitude.

The recently released MS MARCO v2 re-ranking dataset \cite{msmarco16} is the first public test collection large enough to easily reproduce neural IR models.\footnote{Easy in this context means: MS MARCO has enough training samples to successfully train the neural IR models without the need of a bag of tricks \& details applied to the pre-processing \& training regime -- which are often not published in the accompanying papers} The public nature of the dataset also makes it the prime contestant for replicability efforts for neural IR models. 

Nogueira et al. \cite{nogueira2019passage,nogueira2019document} first showed the substantial effectiveness gains for the MS MARCO passage re-ranking using \textit{BERT} \cite{devlin2018bert}, a large pre-trained transformer based model. However, they note the stark trade-off with respect to performance. MacAvaney et al. \cite{macavaney2019} show that by combining \textit{BERT}'s classification label with the output of various neural models exhibits good results for low-training-resource collections. They also show that this comes at a substantial performance cost -- \textit{BERT} taking two orders of magnitude longer than a simple word embedding.

On one hand, the retrieval results achieved with \textit{BERT}'s contextualized encoding are truly impressive, on the other hand, the community should not lose focus of the practicality of their solutions -- requiring fast performance for search. Complementing our argument we present a case study about the performance of different neural IR models and embedding models (Section \ref{sec:case_Study}). We show that using a FastText \cite{bojanowski2017enriching} encoding provides a small trade-off between effectiveness and performance, whereas \textit{BERT} shows a big trade-off in both directions. \textit{BERT} is more than 100 times slower than non-contextualized ranking models. 

The medical computer vision community has already recognized the need for a focus on run time considerations. The medical image analysis benchmark VISCERAL \cite{jimenez2016cloud} included run time measurements of participant solutions on the same hardware. Additionally, computer vision tasks, such as object detection and tracking, often require realtime results \cite{huang2017speed}. Here, iterations over neural network architectures have been focusing on performance \cite{ren2015faster,redmon2016you}. The object detection architectures commonly start with a pre-trained feature extraction model. As Huang et al. \cite{huang2017speed} show, this feature extraction stage can easily be swapped to accommodate different performance-effectiveness needs. We postulate that for neural IR models the time has come to do the same. Neural IR models depend on an encoding layer and recent works \cite{hofstaetter_sigir_2019, macavaney2019, nogueira2019passage} show that the neural IR community has at least 4 different encoding architectures to choose from (basic word embedding, FastText, ELMo, BERT).

The public comparison of results on leaderboards and evaluation campaigns sparks interest and friendly competition among researchers. However, they naturally incentivise a focus on the effectiveness metrics used and other important aspects of IR systems -- for example the latency of a response -- are left aside. The introduction of docker-based submissions of complete retrieval systems makes the comparison of run time metrics feasible: All system can be compared under the same hardware conditions by a third party.

Concretely, we propose to extend the docker-based replicability infrastructure for two additional use cases (Section \ref{sec:scenarios}):

\begin{enumerate}
\item \textbf{Dynamic full system benchmark}\\ We measure the query latency and throughput over a longer realistic period of a full search engine (possibly including a neural  re-ranking component). We envision a scripted "interactive" mode, where the search engine returns results for a single query at a time, giving the benchmark a lot of fidelity in reporting performance statistics.
\item \textbf{Static re-ranking benchmark}\\We measure the (neural) re-ranking components in isolation, providing them with the re-ranking candidate list. This allows for direct comparability of models as all external factors are fixed. This static scenario is very close to the way neural IR re-ranking models are evaluated today, with added timing metrics.
\end{enumerate}

A standardized performance evaluation helps the research community and software engineers building on the research to better understand the trade-offs of different models and the performance requirements that each of them have. It is our understanding that the replicability efforts of our community are not only for \textit{good science}, they are also geared towards the usability of our innovations in systems that people use. We argue that the performance is a major contributor to this goal and therefore worthwhile to study as part of a broader replicability and reproducibility push.

%\newpage
\vspace{-0.2cm}\section{Neural IR Model performance}\label{sec:case_Study}

In the following case study we take a closer look at the training and inference time as well as GPU memory requirements for different neural IR models. Additionally, we compare the time required to re-rank a query with the model's effectiveness.
\vspace{-0.2cm}
\subsection{Neural IR Models}

We conduct our experiments on five neural IR models using a basic Glove \cite{pennington2014glove} word embedding and FastText \cite{bojanowski2017enriching}. Additionally, we evaluate a \textit{BERT} \cite{devlin2018bert} based ranking model. We use the MS MARCO~\cite{msmarco16} passage ranking collection to train and evaluate the models. All models are trained end-to-end and the word representations are fine-tuned. Now, we give a brief overview of the models used with a focus on performance sensitive components:

%\noindent
\textbf{\emph{KNRM}}~\cite{Xiong2017} applies a differentiable soft-histogram (Gaussian kernel functions) on top of the similarity matching matrix of query and document tokens -- summing the interactions by their similarity. The model then learns to weight the different soft-histogram bins. 

%\noindent
\textbf{\emph{CONV-KNRM}}~\cite{Dai2018} extends \textit{KNRM} by adding a Convolutional Neural Network (CNN) layer on top of the word embeddings, enabling word-level n-gram representation learning. \textit{CONV-KNRM} cross-matches n-grams and scores $n^2$ similarity matrices in total. 

%\noindent
\textbf{\emph{MatchPyramid}}~\cite{Pang2016} is a ranking model inspired by deep neural image processing architectures. The model first computes the similarity matching matrix, which is then applied to several stacked CNN layers with dynamic max-pooling to ensure a fixed size output. 

\textbf{\emph{PACRR}}~\cite{hui2017pacrr} applies different sized CNN layers on the match matrix followed by a max pooling of the strongest signals. In contrast to MatchPyramid, the CNNs are only single layered, focusing on different n-gram sizes and single word-to-word interactions are modeled without a CNN.

\textbf{\emph{DUET}}~\cite{mitra2019updated} is a hybrid model applying CNNs to local interactions and single vector representation matching of the query and document. The two paths are combined at the end of the model to form the relevance score. \textit{Note: We employed v2 of the model. We changed the local interaction input to a cosine match matrix -- in line with the other models -- in contrast to the exact matching in the published DUET model. We were not able to reproduce the original exact match results, however the cosine match matrix shows significantly better results than in \cite{mitra2019updated}.}

\textbf{\emph{BERT$_{\textbf{[CLS]}}$}}~\cite{devlin2018bert} differs strongly from the previously described models. It is a multi-purpose transformer based NLP model. We follow the approach from Nogueira et al. \cite{nogueira2019passage} and first concatenate the query and document sequences with the \textit{[SEP]} indicator. Then, we apply a single linear layer on top of the first \textit{[CLS]} token to produce the relevance score. 

\vspace{-0.15cm}
\subsection{Experiment Setup}

In our experiment setup, we largely follow Hofst{\"a}tter et al. \cite{hofstaetter_sigir_2019}. We use PyTorch \cite{pytorch2017} and AllenNLP \cite{Gardner2017AllenNLP} for the neural models and Anserini \cite{Yang2017anserini} to obtain the initial BM25 rankings. The BM25 baseline reaches 0.192 MRR@10 -- as all neural models are significantly better, we omit it in the rest of the paper. We use the Adam optimizer and pairwise margin ranking loss with a learning rate of $1e^{\text{-}3}$ for all non-\textit{BERT} models; for \textit{BERT} we use a rate of $3e^{\text{-}6}$ and the "bert-base-uncased" pre-trained model\footnote{From: \textit{\url{https://github.com/huggingface/pytorch-pretrained-BERT}}}. We train the models with a batch size of 64; for evaluation we use a batch size of 256. We keep the defaults for the model configurations from their respective papers, except for MatchPyramid where we follow the 5-layer configuration from \cite{hofstaetter_sigir_2019}. For the basic word embedding we use a vocabulary with a minimum collection occurrence of 5. The nature of the passage collection means we operate on fairly short text sequences: We clip the passages at 200 and the queries at 20 tokens -- this only removes a modest number of outliers.

In their work, Hofst{\"a}tter et al. \cite{hofstaetter_sigir_2019} evaluate the effectiveness of the models along the re-ranking depth (i.e. how many documents are re-ranked by the neural model) -- they show that a shallow re-ranking depth already saturates most queries. This insight can be employed to tune the performance of re-ranking systems further in the future. In our case study, we keep it simple by reporting the best validation MRR@10 (Mean Reciprocal Rank) results per model. 

We present average timings per batch assuming a batch contains a single query with 256 re-ranking documents. We report timings from already cached batches -- excluding the pre-processing and therefore reducing the considerable negative performance impact of Python as much as possible. We use a benchmark server with NVIDIA GTX 1080 TI (11GB memory) GPUs and Intel Xeon E5-2667 @ 3.20GHz CPUs. Each model is run on a single GPU. %The servers CPU and main memory are never fully utilized.

We caution that the measurements do not reflect production ready implementations -- as we directly measured PyTorch research models and we strongly believe that overall the performance can further be improved by employing more inference optimized runtimes (such as the ONNX runtime\footnote{\textit{\url{https://github.com/microsoft/onnxruntime}}}) and performance optimized support code (for example non-Python code feeding data into the neural network). We would like to kick-start innovation in this direction with our paper.

\begin{table}[t]
    \centering
    \caption{Training performance (Triples/second includes: 2x forward, 1x loss \& backward per triple), Training duration (best validation result after batch count), peak GPU memory requirement as well as the effectiveness score (MRR@10)}
    \label{tab:train_time}
    %\vspace{-0.3cm}
    \setlength\tabcolsep{6pt}
    \begin{tabular}{clrr!{\color{lightgray}\vrule}r!{\color{lightgray}\vrule}r}
       \toprule
       & % emb/fasttext marker
       \multirow{2}{*}{\textbf{Model}} & 
       \textbf{Triples} & 
       \textbf{Batch} & 
       \textbf{Peak} &
       \multirow{2}{*}{\textbf{MRR}} \\
        && \textbf{/ second} &  \textbf{count} & \textbf{Memory}  & \\ \midrule
       \parbox[t]{2mm}{\multirow{5}{*}{\rotatebox[origin=c]{90}{\small\textbf{Word vectors}}}} & 
        \textbf{KNRM}          & 5,200 & 44,000  & 2.16 GB & 0.222  \\
       &\textbf{C-KNRM}        & 1,300 & 98,000  & 2.73 GB & 0.261  \\
       &\textbf{MatchP.}       & 2,900 & 178,000 & 2.30 GB & 0.245  \\
       &\textbf{PACRR}         & 2,900 & 130,000 & 2.21 GB & 0.249  \\
       &\textbf{DUET}          & 1,900 & 146,000 & 2.47 GB & 0.259  \\ 

       \midrule
       \parbox[t]{2mm}{\multirow{5}{*}{\rotatebox[origin=c]{90}{\small\textbf{FastText}}}} & 
        \textbf{KNRM}          & 2,300 & 62,000  & 7.34 GB & 0.231  \\
       &\textbf{C-KNRM}        & 1,000 & 184,000 & 7.81 GB & 0.273  \\
       &\textbf{MatchP.}       & 1,800 & 182,000 & 7.47 GB & 0.254  \\
       &\textbf{PACRR}         & 1,700 & 100,000 & 7.40 GB & 0.257  \\
       &\textbf{DUET}          & 1,600 & 182,000 & 7.46 GB & 0.271  \\ 
       \midrule
       
       &\textbf{BERT$_\textbf{[CLS]}$}    & 33 & 77,500 & 7.68 GB &  0.347  \\ 
        \bottomrule
    %\vspace{-0.4cm}
    \end{tabular}
\end{table}
\begin{table}[t]
    \centering
    %\vspace{-0.2cm}
    \caption{Re-ranking speed (256 documents per query \& batch), peak GPU memory requirement and MRR@10 effectiveness of our evaluated neural IR models.}
    \label{tab:inference_time}
    %\vspace{-0.3cm}
    \setlength\tabcolsep{6pt}
    \begin{tabular}{clrr!{\color{lightgray}\vrule}r!{\color{lightgray}\vrule}r}
       \toprule
       & % emb/fasttext marker
       \multirow{2}{*}{\textbf{Model}} & 
       \textbf{Docs} & 
       \textbf{Time} & 
       \textbf{GPU} &
       \multirow{2}{*}{\textbf{MRR}} \\
        && \textbf{/ second} &  \textbf{/ query} &  \textbf{Memory}  & \\ \midrule
       \parbox[t]{2mm}{\multirow{5}{*}{\rotatebox[origin=c]{90}{\small\textbf{Word vectors}}}} & 
        \textbf{KNRM}          & 48,000 & 5 ms  & 0.84 GB & 0.222  \\
       &\textbf{C-KNRM}        & 12,000 & 21 ms & 0.93 GB & 0.261  \\
       &\textbf{MatchP.}       & 28,000 & 9 ms  & 0.97 GB & 0.245  \\
       &\textbf{PACRR}         & 27,000 & 9 ms  & 0.91 GB & 0.249  \\
       &\textbf{DUET}          & 14,000 & 18 ms & 1.04 GB & 0.259  \\ 

       \midrule
       \parbox[t]{2mm}{\multirow{5}{*}{\rotatebox[origin=c]{90}{\small\textbf{FastText}}}} & 
        \textbf{KNRM}          & 36,000 & 7 ms  & 2.59 GB & 0.231  \\
       &\textbf{C-KNRM}        & 11,000 & 23 ms & 2.68 GB & 0.273  \\
       &\textbf{MatchP.}       & 23,000 & 11 ms & 2.72 GB & 0.254  \\
       &\textbf{PACRR}         & 21,000 & 12 ms & 2.67 GB & 0.257  \\
       &\textbf{DUET}          & 17,000 & 15 ms & 2.68 GB & 0.271  \\ 
       \midrule
       
       &\textbf{BERT$_\textbf{[CLS]}$}    & 130 & 1,970 ms & 7.29 GB &  0.347  \\ 
        \bottomrule
    %\vspace{-0.8cm}
    \end{tabular}
\end{table}

\begin{figure}[t]
    %trim={<left> <lower> <right> <upper>}
   \includegraphics[trim={0.7cm 0.5cm 0.2cm 0.4cm} ,
   width=0.475\textwidth]{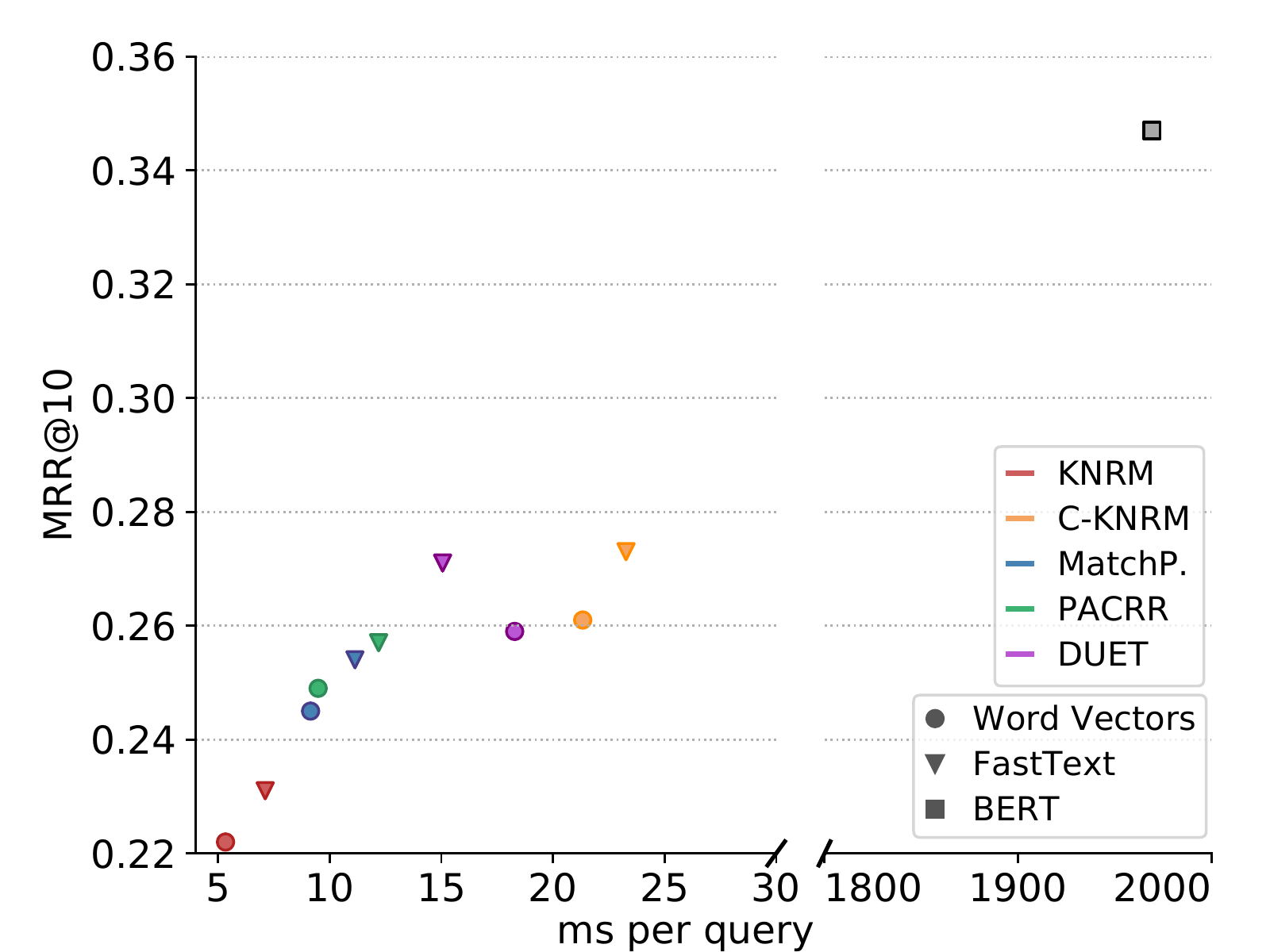}
    \centering
    %\vspace{-0.2cm}
    \caption{A comparison of performance and effectiveness  \textit{Note: the break in the x-axis indicates a large time gap}}
    \label{fig:comparison_plot}
    \vspace{-0.5cm}
\end{figure}

\subsection{Results \& Discussion}

We start our observations with the training of the models as shown in Table \ref{tab:train_time}. The main performance metric is the throughput of triples per second. A triple is a single training sample with a query and one relevant and one non-relevant document. The models are trained with a pairwise ranking loss, which requires two forward and a single backward pass per triple. The batch count is the number of batches with the best validation result. \textit{KNRM} is the fastest to train, it also saturates first. \textit{MatchPyramid} and \textit{PACRR} exhibit similar performance. This is due to their similar architecture components (CNNs applied on a 2D match matrix). In the class of CNNs applied to higher dimensional word representations, \textit{DUET} is slightly faster than \textit{CONV-KNRM}, although \textit{CONV-KNRM} is slightly more effective. In general, FastText vectors improve all models with a modest performance decrease. The peak GPU memory required in the training largely depends on the encoding layer \footnote{We report peak memory usage provided by PyTorch, however we observed that one requires additional GPU memory, FastText \& \textit{BERT} are not trainable on 8 GB. We believe this is due to memory fragmentation. The size of the required headroom remains an open question for future work.}. Fine-tuning the \textit{BERT} model is much slower than all other models. It also is more challenging to fit on a GPU with limited available memory, we employed gradient accumulation to update the weights every 64 samples. We did not observe big performance differences between batch sizes. 

Now we focus on the practically more important aspect: the re-ranking performance of the neural IR models. In Table \ref{tab:inference_time} we report the time that the neural IR models spend to score the provided query-document pairs. The reported time only includes the model computation. This corresponds to benchmark scenario \#2 (Section \ref{sec:scenario2}).

The main observation from the re-ranking performance data in Table \ref{tab:inference_time} is the striking difference between \textit{BERT} and non-\textit{BERT} models. Both the word vector and FastText encodings have a low memory footprint and the level of performance makes them suitable for realtime re-ranking tasks. There are slight trade-offs between the models as depicted in Figure \ref{fig:comparison_plot}. The differences correspond to the training speed, discussed above. However, compared to \textit{BERT}'s performance those differences become marginal. \textit{BERT} offers impressive effectiveness gains at a substantial performance cost. We only evaluate a single \textit{BERT} model, however the performance characteristics should apply to all \textit{BERT}-based models. 

We believe that the practical applicability of \textit{BERT}-based re-ranking models is currently limited to offline scoring or domains where users are willing to accept multiple second delays in their search workflow. Future work will likely focus on the gap between the contextualized and non-contextualized models -- both in terms of performance and effectiveness. Another path is to speed-up \textit{BERT} and other transformer based models, for example with pruning \cite{voita2019analyzing}. Therefore, we argue that it is necessary to provide the replicability infrastructure with tools to take both performance and effectiveness dimensions into account. 

\begin{figure*}[t]
    %trim={<left> <lower> <right> <upper>}
   \includevisio[width=0.9\textwidth]{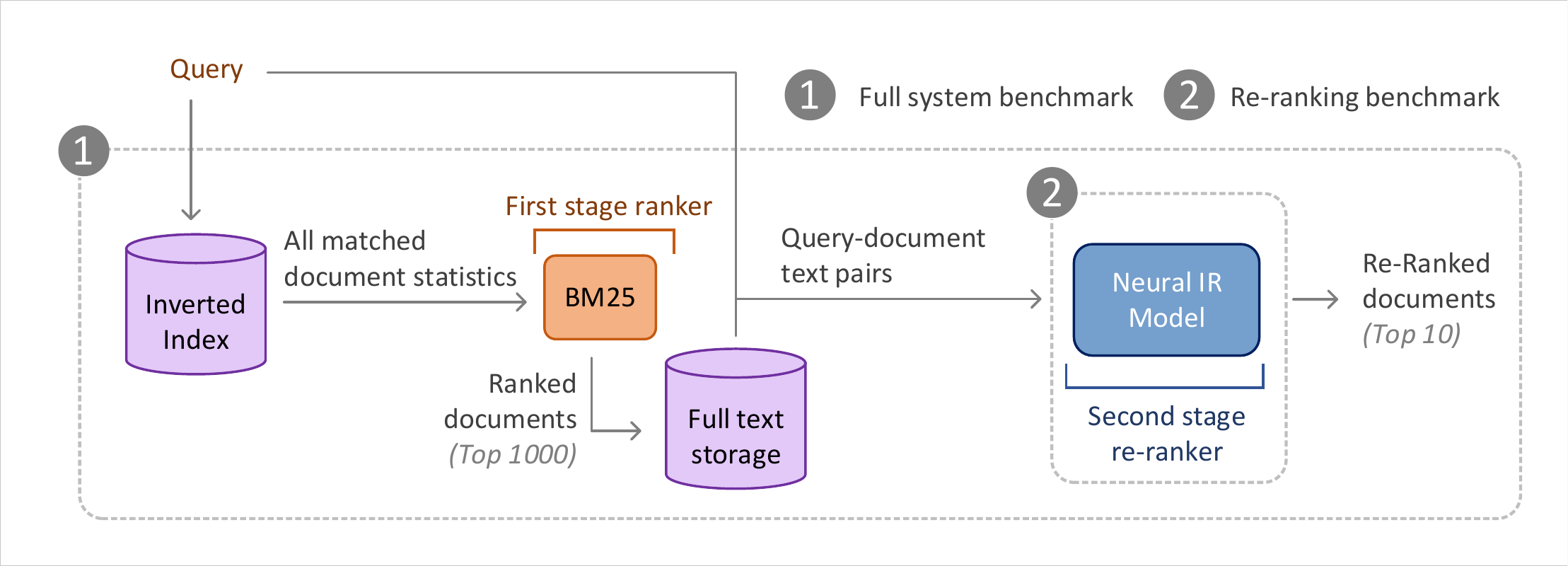}
    \centering
    \caption{A simplified query workflow with re-ranking -- showing the reach of our proposed performance benchmarks}
    \label{fig:scenarios}
    %\vspace{-0.4cm}
\end{figure*}

\section{Benchmark scenarios}\label{sec:scenarios}

Following the observations from the case study above, we propose to systematically measure and report performance metrics as part of all replicability campaigns. Concretely, we propose to extend the OSIRRC docker-based replicability infrastructure for two additional use cases. The different measured components are depicted in Figure \ref{fig:scenarios}.

\subsection{Full System Benchmark}

Currently, most IR evaluation is conducted in batched processes -- working through a set of queries at a time, as we are mostly interested in the effectiveness results. The OSIRRC specifications also contain an optional timing feature for batched retrieval\footnote{See: https://github.com/osirrc/jig}. While we see this as a good first step, we envision a more performance focused benchmark: A scripted "interactive" mode, where the system answers one query at a time. Here, the benchmark decides the load and is able to measure fine grained latency and throughput. 

The scripted "interactive" mode needs as little overhead as possible, like a lightweight HTTP endpoint receiving the query string and returning TREC-formatted results. The execution of the benchmark needs to be conducted on the same hardware, multiple times to reduce noise. 

Although we present a neural IR model case study, we do not limit this benchmark scenario to them -- rather we see it as an opportunity to cover the full range of retrieval methods. For example, we are able to incorporate recall-boosting measures in the first stage retrieval such as \textit{BERT}-based document expansion \cite{nogueira2019document} or query expansion with IR-specific word embeddings \cite{Hofstaetter2019ecir}.

Measuring the latency and throughput over a longer realistic period of a full search engine (with neural IR re-ranking component) touches many previously undeveloped components: Storing neural model input values of indexed documents, generating batches on the fly, or handling concurrency. If a neural IR model is to be deployed in production with a GPU acceleration, the issue of concurrent processing becomes important: We observed that slower models also have a higher GPU utilization, potentially creating a traffic jam on the GPU, that in turn would increase the needed infrastructure cost for the same amount of users.

\subsection{Re-ranking Benchmark}\label{sec:scenario2}

The neural IR field is receiving considerable attention and has a growing community. In our opinion, the community is in need of a more structured evaluation -- both for performance and effectiveness. We now propose a benchmark, which aims to deliver on both dimensions.

The re-ranking benchmark focuses on the innermost component of neural IR models: the scoring of query-document tuples. We provide the re-ranking candidate list and the neural IR model scores the tuples. Many of the existing neural IR models follow this pattern and can therefore easily be swapped and compared with each other -- also on public leaderboards, such as the MS MARCO leaderboard. This static scenario provides a coherent way of evaluating neural IR re-ranking models. It helps to mitigate differences in the setup of various research groups.
\section{Conclusion}

The OSIRRC docker-based IR replicability infrastructure presents an opportunity to incorporate performance benchmarks. As an example for the need of a broader view of the community, we show in a case study the trade-off between performance and effectiveness of neural IR models, especially for recent \textit{BERT} based models. As a result, we propose two different performance-focused benchmarks to be incorporated in the infrastructure going forward. We look forward to working with the community on these issues.

\bibliographystyle{ACM-Reference-Format}
\bibliography{bibliography}
\balance
\end{document}